# Aqueous Alteration Studies on Mukundpura (MK) Carbonaceous Chondrite using FTIR, TGA and Raman spectroscopy and its CM classification

A. Dixit[1,*] R. P. Tripathi[2], Sudhanshu Kumar[3], Mohd. Azaj Ansari[3], K Sreenivas[3]

[1]Department of Physics & Center for Solar Energy, Indian Institute of Technology Jodhpur, 342037, India
[2]78, BGT Extension, Jodhpur 342005, India
[3]Department of Physics and Astrophysics, University of Delhi, Delhi, 110007, India
*ambesh@iitj.ac.in

**Abstract**:

FTIR measurement on MK immediately after its fall, shows a unique doublet around 10 μm, significantly different from many ordinary CM2 chondrites, where only a singlet around 10 μm is observed. Also a very faint 11.2 μm feature in MK indicates the absence of anhydrous silicates, olivine and thus complete serpentinization of anhydrous silicates due to severe aqueous alteration on parent body.  Raman studies show a low peak metamorphic temperature around 0 °C and consistent with the absence of peak corresponding to tochilinite in FTIR spectrum, which forms at higher temperature. The first thermogravimetric measurement was carried out within 24 hrs. of MK fall, showing 10% weight loss in 400-770 °C range and is consistent with TGA on another MK fragment of same batch after 30 months, confirming  no environmental  impact on the water bound to  hydrated clay. This large weight loss also rules out any post aqueous alteration thermal event suffered by MK and signify the presence of hydrated clay. The measured ratio of MgO/FeO is about 0.56, and sulfur weight is 3.4 %. Recently, based on only aqueous alteration Potin et al. (Potin et al. 2020), classified MK as CM1, implying MK should be utmost altered CM, whereas Ray et al. (Ray et al 2018), as equivalent to Paris like (CM2.7), least aqueous altered.  However, if we combine the observed aqueous alteration, MgO/FeO weight ratio, and sulfur weight % together will provide a more comprehensive understanding for MK, and thus, we  classify it as CM2.3.insted CM1 or CM(2.7-2.9).

**Introduction:**

Carbonaceous chondrites are very complex rocks that carry important information about the processes and conditions in the early nebula and the parent bodies, including the active processes in the solar nebula and the parent bodies. The study of the mineral altered and formed by the aqueous alteration processes within meteorite parent bodies is of fundamental importance in various aspects of solid matter in the early solar system. Several aspects of aqueous alteration of carbonaceous chondrite in detail were reviewed by Bischoff et al.,1998(Bischoff 1998). An extensive study is reported on CM, and CI chondrites in the literature, based on a variety of techniques because it is believed that all the alteration features appearing in CM and CI chondrites are consequences of processes that have taken place on the parent body ((King et al. 2015) (Bates et al. 2020)(Jilly-Rehak et al. 2017)(Beck et al. 2010) (Morlok 2013))

Aqueous alteration of CM type chondrites results in the formation of many hydrous silicates (especially serpentine group minerals, e.g., cronstedite, i.e., Fe -serpentine, antigorite, etc.) and sulfur-containing phases like tochilinite. The simultaneous aqueous alteration also results in the formation of various organic compounds both aliphatic, aromatic, and polyaromatic hydrocarbons and their alkylation products (Sephton 2002). The degree of aqueous alteration varies from meteorite to meteorite. Even a different class of the same meteorite may show different degrees of alteration, e.g., among CM chondrites, Paris Carbonaceous chondrite is considered the least aqueously altered (Hewins et al. 2014)(Martins et al. 2015). Petrological, geochemical, chemical, and many spectral parameters are related to the degree of aqueous alteration, so these studies are frequently used to estimate the degree of aqueous alteration (Takir et al. 2013). It is worth mentioning that the temperature range in which alteration took place for most of the CM2 chondrite on the parent body crudely falls between 0 to 100 °C (Cobb and Pudritz 2014). However, it is well known that many meteorites also suffer heating due to radiogenic and shock-related metamorphism which is the main cause of dehydration (i.e., post hydroxylation) of phyllosilicates. We refer study carried out on the fragments of Jbilet Winselwan for CM type CC (King et al. , 2019) for such aqueous alteration. An extensive survey of the literature shows that most of the found CCs suffered during their fall considerably while exposing to the environment before reaching the research laboratories for measurements and showed valence fluctuation of iron as well (Garenne et al. 2019). But Mukundpura carbonaceous chondrite (hereafter we will refer to this as MK) is unique in the sense that it was recovered after its fall within six hours, and fragments of this meteorite were used in the present study. The preliminary studies were reported earlier by our group (Tripathi et al. 2018). Thus, MK is the freshest chondrite ever studied because of the quick recovery and proper handling.

MK fall occurred in dry alluvium farmland in Mukundpura village near Jaipur, Rajasthan (INDIA) on early morning of 6$^{th}$ June 2017. Month of June is the hot summer season in Jaipur, Rajasthan, and the temperature of Jaipur was 45 °C on the day of MK fall with no humidity, and the air was completely dry. The rock broke into several pieces on impact and made a shallow pit. The major fraction of this rock was collected by the Geological Survey of India (GSI). A small fragment of this meteorite was picked by one of the authors (RPT), happens to reside close to the fall site, and personally collects the fragments within six hours of the fall. He took special precautions to avoid terrestrial contamination during collection and storage. The MK fragments were immediately wrapped in aluminum foil, and the sample was stored in a glass container. This fragment was black and gave a pungent smell of sulfur. The first thermogravimetric analysis (TGA) and room-temperature Mössbauer spectroscopic investigations were carried out within a week after the collection of this meteorite. Room temperature Mössbauer spectrum showed only two doublets corresponding to iron in $Fe^{2+}$ and $Fe^{3+}$ state in phyllosilicates, but no characteristic doublet corresponding to iron in olivine (i.e., iron-containing anhydrous silicates) or iron in other minerals was present. Also, no sextet was present in the room temperature Mössbauer spectrum. Mössbauer pattern observed closely resembled those observed by Burns and Fisher for Cold Bokkeveld (CM2) meteorite at room temperature (Burns and Fisher 1994). Thus, it was inferred that MK is also a CM2 chondrite and has experienced extreme serpentinization. This preliminary study was reported by Tripathi et al. (2018)

The detailed study on the sample collected by (RPT) was carried out by Rudraswamy et al. (Rudraswamy et al. 2019) and Dixit et al. (Dixit et al. 2019), and supported the findings of Tripathi et al. (Tripathi et al.

2018) that this meteorite is CM2 type and suffered a higher degree of alteration. Rudraswamy et al. (2018) also reported that MK has a low abundance of olivine and a high abundance of serpentine minerals (Rudraswami et al. 2018). The ratio of FeO/SiO$_2$ =1.05 in poorly characterized phases (PCP) is consistent with that observed in other CM2 chondrites. They also found a meager presence of chondrules (7%) and inferred that most of the chondrules formed initially have been significantly altered or dissolved by alterations on the parent body (Zolensky et al. 1993). In present work FTIR spectroscopy, we show that FTIR pattern of MK is very much different around 10 µm feature. Further, Raman spectroscopy is used to get a broad idea about the temperature on parent body and for the first time we report Raman spectrum at liquid nitrogen temperature and at different temperature while rewarming. Thermogravimetric (TGA) analysis is used to determine water content and hydrogen wt % to show degree of aqueous alteration. By combining the observed aqueous alteration, MgO/FeO weight ratio, and sulfur weight% indicate that it should be classified as CM2.3 instead Ray et al. (Ray et al. 2018) earlier classified it as CM2.7 or recently Potin et al. (Potin et al.2020) classified as CM1.

**Experimental details:**

FTIR spectra were recorded for the powder sample using a Thermosphere (Nicoleta iS50)FT-IR spectrometer. For better resolution, the FTIR spectra were recorded with 64 scans in ATR mode, which may extend the low frequency up to 100 cm$^{-1}$. Raman spectra were collected for the powder sample using a Renishaw (Invia ll) spectrometer equipped with a 514 nm laser operating at 1.0 mW, providing 1 µm$^2$ spot size. The optical arrangements include a triple monochromator fitted with a NEXT filter and provide high resolution (0.5 cm$^{-1}$) measurement with a spot size of 1 µm$^2$ and extends the low frequency up to 50 cm$^{-1}$.

The carbon, hydrogen, nitrogen, and sulfur (CHNS) analysis is carried out using the Elementar Vario EL cube CHNS analyzer. In the case of TGA, two different powder fractions were used to estimate weight loss on two other systems with a long time gap of about 30 months. The first TGA measurement was carried out within 24 hours of this meteorite fall using a Perkin Elmer TGA system at Malviya National

Institute of Technology Jaipur (India). About 20 mg MK sample was used for heating in a nitrogen atmosphere with a 20ml/min flow rate. A total of 44160 data points (very high resolution) were collected. Another fraction was studied around 30 months after MK's fall using Oil India Corporation limited facility situated at Faridabad, India. The system used was NETZSCHSTA 449FI TGA analyzer. The main aim to repeat this measurement was to determine if there was any difference in weight loss in the range 440-770 °C to compare with our first result studied after 24hrs.

## 3. Results and Discussion:

### 3.1 FTIR and Aqueous alteration

FTIR measurements on Mukundpura meteorite are shown in Fig 1 (a & b). The peaks are marked for easy identification. The feature at around 2.85 µm (about 3434 cm$^{-1}$) signifies instead Mg-OH stretching mode indicating phyllosilicates' presence. Interestingly, the 10 µm feature in FTIR is showing a doublet instead of a sharp peak. The vibrational modes around 10 µm, i.e., 9.43 µm (1054 cm$^{-1}$) and 10.4 µm (959 cm$^{-1}$) correspond to the Si-O bond's stretching, substantiating the XRD observations of phyllosilicates phases. Interestingly, these modes are a bit different from Kings' report (Bates et al. 2020; King et al. 2015), and the lower shift of such modes is because of the presence of serpentine. The higher wavelength mode corresponds to cronstedtite present in Mukundpura meteorite. In contrast, the lower wavelength mode may be either due to Mg-rich serpentine or Si-O stretching in silicon sheets.

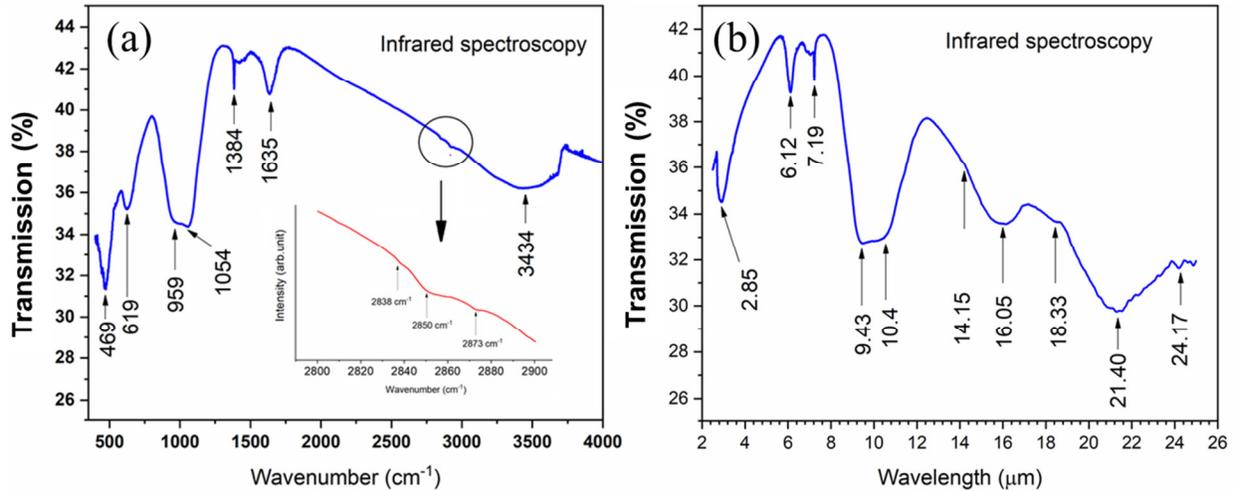

Figure 1: Fourier transform infrared (FTIR) measurements showing % Transmission against (a) wavenumber and (b) wavelength on Mukundpura meteorite.

Further, Kings et al. (2019) reported on the presence of sharp 10 μm features in FITR spectra in Jbilet Winselwan, Murchinson, and Santa Cruz meteorites only (King et al. 2015). Similar observations are made on Cold Bokkeveld, Murchinson, Murrey (CM chondrites), and in Ivuna, Orgeeil, and Aliad (CI) type meteorites by Beck et al. and is a unique feature of this meteorite(Beck et al. 2010). The doublets were also expected because samples of terrestrial Fe cronstedtite or Mg-serpentine show unique 10 μm feature, where instead of a sharp peak at 10 μm, two distinct peaks for Si-Si stanching's are present in the FTIR spectrum. This is the unique difference observed in MK. Though here, our objective is to show the difference between MK and other studies. But at the same time, it also opens a question that if this difference (presence of two peaks around 10 μm feature in MK) is related to the different parent body. This interesting feature in MK, observed in the present study, differs from the recent independent FTIR study of MK by Potin et al. (Potin et al. 2020), reported one broad peak around 10 μm only.

Apart from this, there is clear evidence of a 6.12 μm feature, showing the presence of a water molecule, and a 7.19 μm feature supports the presence of carbonate in MK. The most exciting result is that it shows several features between 2500 cm$^{-1}$ - 300cm$^{-1}$ (see inset). These peaks correspond to organic CH-CH and CH$_2$ -CH$_2$ stretching. However, the most interestingly sharp 11.2 μm feature, corresponding to olivine, is absent in MK FTIR spectrum, Fig 3. The presence of water molecules and hydrocarbon bands suggest that it has experienced a high degree of aqueous alteration without experiencing high temperature.

**3.2 a Raman Spectroscopy and broad estimation of temperature at parent body**

Raman spectroscopy is one of the very important non-destructive measurement techniques to characterize carbon-based materials. MK meteorite is a rare carbonous meteorite and clearly showing shows two vibrational modes at 1350 cm$^{-1}$ (~ 7.4 µm) and 1600 cm$^{-1}$ (6.3 µm), which correspond to D and G bands of graphite (or carbon-based organic materials), Fig 2. Interestingly, these modes are also clearly discernible in FTIR spectra (Fig 1), showing the evidence of carbonaceous content in Mukundpura meteorite. D band presence, i.e., mode at 1350 cm-1 (~ 7.4 µm), is attributed to the disorder in graphene, i.e., sp$^2$-hybridized carbon matrix, giving rise to the resonance at this wavenumber. Further, the G band, i.e., vibration mode at1600 cm$^{-1}$ (6.3 µm) is attributed to C-C stretching and is usually observed in sp$^2$ hybridized carbon materials. These modes are very wide in nature, suggesting the contribution of distorted carbon or graphitic structure in organic materials. The broad 1600 cm$^{-1}$ G peak can be fitted with two peaks at 1580 and 1620 cm$^{-1}$, which corresponds to G and D* vibrational modes, respectively. This broadness is usually attributed to the localized interaction of defect mode with graphene phonon modes resulting in two such vibrational modes. Temperature and field-dependent magnetic and temperature-dependent Mössbauer studies showed the glassy spin nature of MK meteorite, suggesting the random localized distribution of magnetic ions(Dixit et al. 2019). Further, XRD measurements showed broad peaks, supporting nanocrystallites' presence in powder forms, which is substantiated by SEM studies. Thus, observation of such broad D and G bands in MK meteorite sample suggests the presence of large nanostructured carbonaceous material as well.

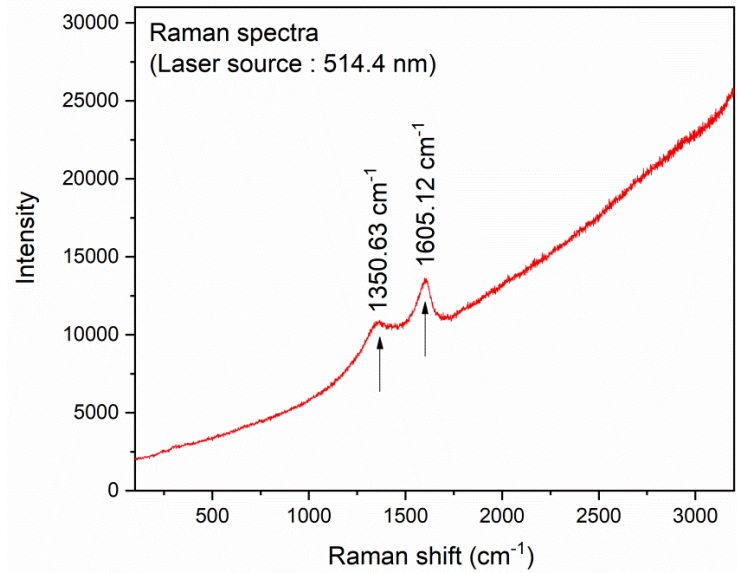

Figure 2: Room temperature Raman spectrum for Mukundpura meteorite, showing two clear modes at 1350 cm$^{-1}$ and 1600 cm$^{-1}$

Further, Raman data can be used as a geothermometer, and for the same, following Homma et al. scheme (Homma et al. 2000), we fitted four peaks to the Raman data, and the fitted parameters are summarized in Table 1.

Table 1. Parameters from the peak fit of the room temperature Raman data

| Peak position (cm$^{-1}$) | Peak assignment (Homma et al. Scheme) | Error in Peak position (cm$^{-1}$) | Full-width half maxima (FWHM) | Relative Intensity |
|---|---|---|---|---|
| 1236.92 | D4 | 10.18 | 166.16 | 328.58 |
| 1350 | D1 | --- | 154.13 | 2259.98 |
| 1500 | D3 | --- | 184.06 | 875.89 |
| 1600 | GL | --- | 76.19 | 3094.72 |

The assignment of these fitted peaks is according to the scheme proposed by Homma et al. (Homma et al. 2000), and the full-width half maxima of D1 peak is used to estimate the peak metamorphic temperature (PMT) as

$$PMT(°C) = -6.9 \times \Gamma_{D1}(cm^{-1}) + 1054.4 \ (R^2 = 0.83)$$

Where $\Gamma_{D1}$ is FWHM for D1 peak in $cm^{-1}$. This equation applies for 0 °C to 586 °C temperature range thermometer. The estimated PMT is nearly -9.09 °C, which is very close to 0 °C Though this is a crude estimate but nevertheless it can be safely concluded MK experienced low temperature alteration . This is consistent with the absence of peak corresponding to mineral tochilinite in FTIR spectrum and in room temperature Mössbauer spectrum of MK ( mineral which form at higher temperature, Vacher et al. (Vacher et al. 2019)). This is the first analysis on MK (as per authors' knowledge) on PMT using Raman data, which is used to understand the thermal metamorphism of carbonaceous chondritic meteorite. The 0 °C ( or relatively low temperature) PMT value for MK meteorite suggests that it has gone through the least graphitization and least affected thermally.

**3.2 b Low-temperature Raman Spectroscopic measurement**

For the first time, we cool MK sample at liquid nitrogen temperature to see how D and G characteristic vibrational modes change (that is, any change occurs in C-C starching mode due to cooling.) Further, we have heated at room temperature and see if this process is reversible or not. In Fig 3, we have shown the Raman spectra at different temperatures under cooling until -190 °C and under warming until 150 °C.

The most exciting finding is that there is a considerable shift in the unfitted spectrum. The low wavenumber modes correspond to other phases present in MK. There is no shift in Raman modes while warming MK sample. However, from the visual inspection of Fig 3, it can be inferred that the area under D band is drastically decreasing, and around 150 °C area under D band significantly reduces to a minimal area. These are only preliminary investigations, and more studies are needed to understand the significance of these observations. But it will be interesting to note that D band intensity decreases drastically around 100 °C, which is believed as the temperature where most meteorites aqueously altered.

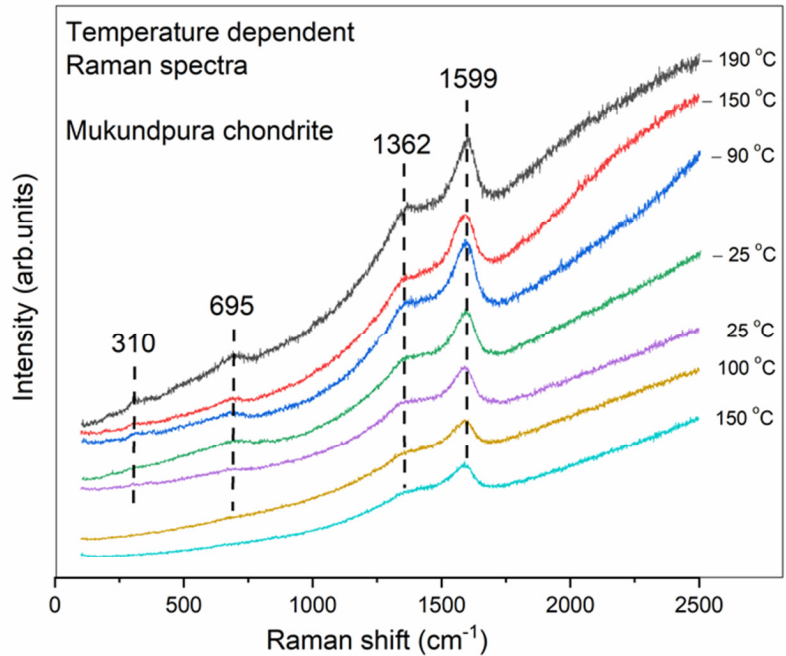

Figure 3: Temperature dependent Raman spectrum for Mukundpura meteorite form room temperature i.e. 25 °C to -190 °C (under cooling) and till 150 °C (under warming).

**3.5 Thermogravimetric analysis (TGA)**

The aqueous alteration of anhydrous magnesium silicates to produce phyllosilicates (mainly serpentine group minerals) is a major process in the formation of CM and CI chondrites. It has also been recognized that reverse dehydration also occurs in a number of meteorites due to metamorphism. An extensive literature survey shows that TGA will show different weight loss in the temperature range 200-770 °C for meteorites, which suffered shock metamorphism or for meteorites escaped shock metamorphism. In the former, relatively less weight loss is observed in comparison to that of later meteorite.

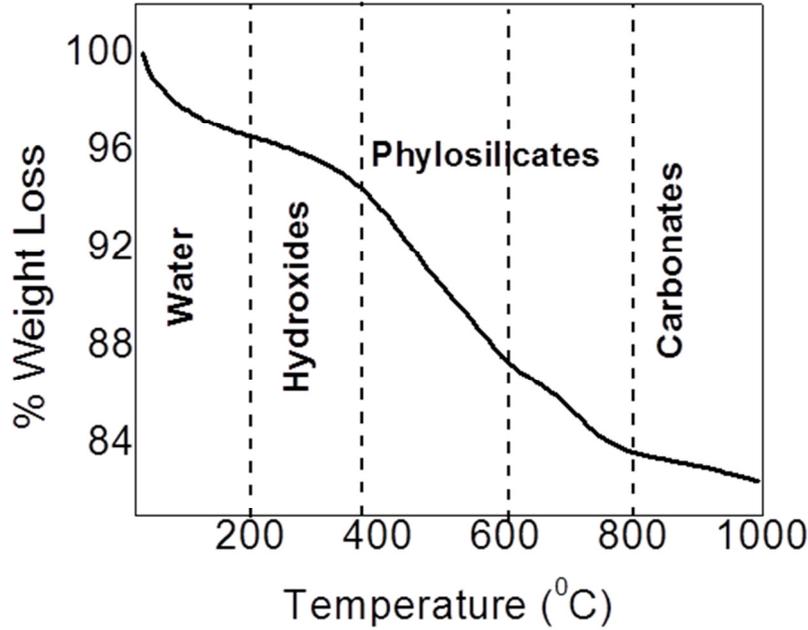

Figure 4. Thermogravimetric analysis of MK meteorite sample after 30 months, showing percent weight loss against temperature.

In the present investigation, we report the relative comparison of TGA measurement which was carried out on pristine sample within a day immediately after the fall under nitrogen ambient, discussed by Tripathi et al. (Tripathi et al. 2018) with on another fraction of same MK sample after thirty months, which is shown in Fig 4. The motive behind these two measurement was to compare if any major difference is reflected in weight loss due to the lapse of time specially in the temperature range 400-770 °C, where heating releases hydroxyl group (OH$^-$) group from phyllosilicates. We summarized the mass loss (wt %) occurs between temperature 25-1000 °C in Table 2 for both samples.

Table -2 Mass loss (wt%) in different temperature range during TGA for Mukundpura Meteorite. Here 24 hrs means TGA measurement carried out within 24 hrs of fall and 30 months TGA taken after the lapse of thirty months.

| MK Sample | Total mass loss (%) | 25 – 200 °C | 200 – 400 °C | 400 – 770 °C | H$_2$O | H$_2$ (wt%) |
|---|---|---|---|---|---|---|
| MK | 16.46 | 2 | 2.03 | 9.84 | 11.87 | 1.37 |
| MK-30 | 17.59 | 3.4 | 2.78 | 9.83 | 12.61 | 1.37 |
| Average | 17.05 | 2.7 | 2.40 | 9.84 | 12.24 | 1.37 |

Here, MK is meteorite sample on which TGA is carried out within 24 hrs after fall, and MK-30 is MK meteorite sample on which TGA is carried out after 30 months

We can divide the entire temperature range into different regions related to dehydration of molecular water (25-200 °C), the dehydration and dihydroxylation of Fe (oxy) hydroxides (200-400 °C), dihydroxylation of phyllosilicates (in the case MK, phyllosilicates are mainly iron rich cronstedtite, and Mg rich serpentine ) in the range (400 - 770 °C) and between (770 - 1000 °C) weight loss is due to decomposition of carbonate minerals and release of carbon dioxide. This sequence which we have discussed is similar as used by Kings et al. (Kings et al. 2019).We also summarized the probable sources of weight loss in different temperature regions in Table 3 for MK samples, examined within 24 hrs and after 30 months..

Table 3: The different temperature regions together with the respective percent loss for pristine MK i.e. within 24 hrs and after 30 months MK samples

| Temperature regions | | Sample | Weight loss (%) | Respective weight loss source |
|---|---|---|---|---|
| RT - 200 °C | | MK | 2.0 | Molecular water |
| | | MK-30 | 3.4 | |
| 200 – 380 °C | | MK | 2.0 | Hydroxides/ oxyhydroxide |
| | | MK-30 | 3.18 | |
| 380-770 °C | 380 – 650 °C | MK | 10.0 | Phyllosilicates (Fe and Mg based serpentines) |
| | 650 - 770 °C | MK-30 | 10.5 | |
| 770 – 1000 °C | | MK | 1.54 | Carbonates |
| | | MK-30 | 1.45 | |

Here RT is room temperature, MK is meteorite sample on which TGA is carried out within 24 hrs after fall, and MK-30 is MK meteorite sample on which TGA is carried out after 30 months.

As we have already mentioned that the temperature on the day of fall and next day was more than 45 °C with no humidity, in such climatic condition it is unlikely that meteorite could have absorbed any significant amount of terrestrial water within six hrs. before it is collected. The first TGA measurement was carried out within 24 hrs. of its fall and whatever weight loss observed for MK in temperature range 25-200 °C (in present case it is 2%) must be due to water already present in MK matrix. This is the first measurement in which we can confidently say that how much extraterrestrial molecular water must be present in CM2 type meteorites. Beck et al. (Beck et al. 2014) have studied water contained in large number of CM type meteorite but in their study most of the samples were find and only few samples were under fall category. The studied meteorites by Beck et al. (Beck et al. 2014) also have different degree of weathering. Apart from this, these meteorites were not studied immediately after their fall so in the range 25-200 °C.. Therefore in most of the studies while estimating water concertation, weight loss observed in this range 25 - 200 °C is excluded while estimating hydrogen%. Beck et al. (Beck et al. 2014) in differential thermogravimetric analysis(DTGA) obtained two intense peaks in most of the samples, corresponding to absorbed water and mesopore water in the range 25-200 °C, but in case of MK, there is only one hump in this temperature range, (Tripathi et al. 2018), at temperature below 100 °C This suggest that there is only one source of water in MK meteorite sample. The appreciable amount of water acquired by the parent body and there is no way that it is derived from terrestrial environment. Weight loss observed for MK is more representative for extraterrestrial molecular water In temperature range between 400-700 °C in 24 hrs sample, two humps at around 600 °C and 750 °C are present. This can be understood as initially from 380 – 600 °C iron rich serpentine minerals are dehydrated whereas from 600 – 750 °C mainly magnesium based serpentine group minerals are dehydrated. If we compare the 30 months TGA with that of collected within 24 hrs, Fig 4, the weight loss is almost identical in temperature range between 400 -770 °C indicating that considered MK sample for present study has not suffered any weathering. There is some variation in other temperature range most probably due to inhomogeneity of distribution of water and other phases. Major weight loss in MK meteorite has occurred in this temperature range due to the dehydration of hydrated phyllosilicate, which is around ~ 10% of the total weight. Further, the total weight loss is app.17%, signifying that it is a CM2 like meteorite, which escaped the metamorphism. Here it will be worth mentioning that CI chondrites show much larger weight loss in comparison to CI chondrites (Bates et al. 2020; King et al. 2015).

We have calculated hydrogen content from the TGA of pristine sample (i.e. using fresh sample with 24 hrs of the fall) using procedure adopted by Kings et al. (King et al. 2019), and summarized in Table-2. We obtained large concentration of hydrogen, further indicating appreciable aqueous alteration.

Simultaneously, we also carried out C, H, N, and S elemental analysis and the recorded fractional percentage is about 2.100%, 0.996% 0.910%, and 3.409%, respectively. The difference in two measurement can be understood that in TGA one always obtains higher value because our assumption that mass loss between 200-770 °C is because of $H_2O$. Though, this is a good approximation but decomposition of some minor phases can also contribute and overlap with phyllosilicates (Alexander et al. 2013) and so it may be the probable upper limit. On the other hand using CHNS analyzer, it is possible to under estimate hydrogen due to very small sample size. So we think average of these two measurements i.e., (1.15wt %) can be considered as the hydrogen content in MK. Kings et al (King et al. 2019) compiled various plots using data from various sources and in a figure (figure 13, King et al 2019) the amount of bulk Hydrogen against phyllosilicates is summarized. The amount of H (~ 1.15%) in MK meteorite corresponds to the presence of 80- 85% phyllosilicate (hydrous) silicates, if estimated from Kings et al. work (King et al. 2019). This is consistent with those reported by Rudraswamy et al (Rudraswamy et al. 2019). This suggests that our estimation of hydrogen is in well acceptable range. In another plot (Fig 12, King et al. 2019)), magnetite concentration is plotted with hydrous phyllosilicates present in various meteorites using data available in the literature. If we assume that this correlation exists then the concentration of magnetite should not exceed ~ 3% in Mukundpura meteorite. This concentration is beyond the detection limit (iron % present in magnetite will be even very much less than 3%) to detect magnetite in Mössbauer spectroscopy. Absence of magnetite sextet in Mössbauer spectra suggests our estimation of magnetite using Kings et al (2019) relation is acceptable. But such small concentration can be sensed in magnetic measurements as reported by Dixit et al (2019).

Beck et al.(Beck et al. 2014) have argued that weight loss between the range 200-770 °C has some correlation with petrographic grade (PG) of meteorite. In Table -4 we have listed weight loss and petrographic grade of some meteorites with respective references.

Table 4. Total weight loss/weight loss between 25 – 1000 °C and 200 – 770 °C together with petrographic grade (PG) of some representative meteorites

| Meteorite | Weight loss % (25 – 1000 °C) | Weight loss % (200 – 770 °C) | PG | Reference |
|---|---|---|---|---|
| ALH83100 | 16.9 | 14.6 | 2.1 | Beck et al. 2014 |
| LEW87022 | 16.1 | 12.8 | 2.3 | Beck et al. 2014 |
| Jubilet-Wilsenwal* | 16.5 | 9.7 | > 4 <7 | Kings et al. 2019 |
| Murchinson | 15.4 | 12.7 | 2.5 | Beck et al. 2014 |
| QUE97990 | 15.0 | 11.0 | 2.6 | Rubin A E, 2007 |

*Jubilet-Wilsenwal suffered metamorphism.

If we compare weight loss given in Table-2 and Table-4, then (a) petrographic grade of MK should be less than four, and (b) it has not suffered any metamorphism. In earlier study Ray and Shukla (2018) suggested MK meteorite is analogous to Paris meteorite, which has petrographic grade (PG) of 2.9. Further, Ray and Shukla (2018) also suggested that it has suffered moderate aqueous alteration; our results conclusively differ or even negate both these arguments from Ray and Shukla (2018). To get idea about petrographic grade of MK, it is worth to note that MgO/FeO ratio in CM chondrites is expected to increase with degree of aqueous alteration. This ratio lies between (a) *0.35-0.42* for CM having *PG > 2.7*, (b) *0.39-0.46* for *PG 2.6 – 2.4*, and (c) *>0.65* for *PG < 2 - 3* (Zolansky et al1993,Hewins et al 2014,Rubin 2015,Lee et al.2016). Rudraswamy et al have already reported that this ratio is around 0.56 for MK meteorite. This rules out that MK is PG grade 2.1. Further, Zolansky et al (Zolansky et al. 2013) and Rubin et al. (Rubin et al. 2015) established that sulfur abundance is higher in least altered CM chondrites 4 wt % in Paris with PG (2.9 - 2.7), 2-3 weight % Murchinson (2.5) Murray (2.5-2.4) ,Y791198 (2.4 PG) , 2 wt% in Nogyoya (2.2). In present case, MK seems a unique meteorite, which on the one hand is highly aqueous altered and also large ~ 3.4% sulfur weight% . This seems like a paradox, while TGA and MgO/FeO ratio suggest its PG must be less than 2.4, and sulfur content suggests its PG should be more than 2.2; thus, the only possibility left is that MK is CM(2.3). This classification is more convincing as carried out on fresh sample and also based on not only aqueous alteration but also on elemental compositions contribution analyzed using petrographic grade of MK and sulfur weight. This differs from the recent studies of MK as CM1 by Potin et al (Potin et al. 2020), which is based on only aqueous alteration and also differ from Ray et al (Ray et al. 2018), classifying it as Paris like i.e. CM2.7. These classifications are based on aqueous alteration and elemental considerations are not considered simultaneously. Thus, it becomes essential to consider the other factors such as petrographic grade, sulfur and hydrogen content in the present analysis for classifying it more convincingly. The present analysis based on aqueous alteration, PG, and sulfur weight in present study provides a more comprehensive understanding of MK and classifies it as CM2.3

**Conclusion:**

We present studies on the relatively fresh sample of MK to avoid any terrestrial contamination. The Raman spectrum in MK CM2 chondrite indicates that aqueous alteration must have taken place at relatively very low temperatures ~ 0 °C. For the first time, we studied the characteristics of D and G bands and also studied the impact of temperature variation on these vibrational modes. It is observed that D band intensity drastically decreases when we heat sample from liquid nitrogen to 150 °C. We substantiated the observed aqueous alteration in MK using Raman spectroscopic and thermogravimetric measurements. Further, combining the degree of aqueous alteration, hydrogen content, sulfur weight, and petrographic grade analysis classify MK as CM2.3.

**Acknowledgement:**

Authors acknowledge Professor N. Bhandari for his fruitful discussion and help in carrying out some of the measurements on MK sample.

**Conflict of Interest:**

Authors declare that there is no conflict of interest.